%% file: main.tex
\documentclass[10pt,conference]{IEEEtran}

\usepackage{amsmath,amsfonts}
\DeclareMathOperator*{\argmax}{arg\,max}

\usepackage{hyperref}
\usepackage{subcaption}
\newsavebox{\mybox}

\input{header}

\begin{document}

\title{Information-Theoretic Testing and Debugging of Fairness Defects in Deep Neural Networks}

\author{
\IEEEauthorblockN{Verya Monjezi}
\IEEEauthorblockA{
vmonjezi@miners.utep.edu \\
\textit{University of Texas at El Paso}}
\\
\IEEEauthorblockN{Gang Tan}
\IEEEauthorblockA{
gtan@psu.edu \\
\textit{Pennsylvania State University}}
\and
\IEEEauthorblockN{Ashutosh Trivedi}
\IEEEauthorblockA{
ashutosh.trivedi@colorado.edu\\
\textit{University of Colorado Boulder}}
\\
\IEEEauthorblockN{Saeid Tizpaz-Niari}
\IEEEauthorblockA{
saeid@utep.edu \\
\textit{University of Texas at El Paso}}
}

\IEEEtitleabstractindextext{
\begin{abstract}
The deep feedforward neural networks (DNNs) are increasingly deployed in socioeconomic critical decision support software systems. 
DNNs are exceptionally good at finding minimal, sufficient statistical patterns within their training data. 
Consequently, DNNs may learn to encode decisions---amplifying existing biases or introducing new ones---that may disadvantage protected individuals/groups and may stand to violate legal protections. 
While the existing search based software testing approaches have been effective in discovering fairness defects, they do not supplement these defects with debugging aids---such as severity and causal explanations---crucial to help developers triage and decide on the next course of action.
Can we measure the severity of fairness defects in DNNs? 
Are these defects symptomatic of improper training or they merely reflect biases present in the training data? 
To answer such questions, we present \toolname{}: an information-theoretic testing and debugging framework to discover and localize fairness defects in DNNs.

The key goal of \toolname{} is to assist software developers in triaging fairness defects by ordering them by their severity. 
Towards this goal, we quantify fairness in terms of protected information (in bits) used in decision making. 
A quantitative view of fairness defects not only helps in ordering these defects, our empirical evaluation shows that it improves the search efficiency due to resulting smoothness of the search space. 
Guided by the quantitative fairness, we present a causal debugging framework to localize inadequately trained layers and neurons responsible for fairness defects.
Our experiments over ten DNNs, developed for socially critical tasks,
show that \toolname{} efficiently characterizes the amounts of discrimination,
effectively generates discriminatory instances (vis-a-vis the
state-of-the-art techniques), and localizes layers/neurons with significant biases.
\end{abstract}

}

\maketitle
\IEEEdisplaynontitleabstractindextext

\input{sections/introduction}

\input{sections/background}

\input{sections/overview}

\input{sections/definition}

\input{sections/approach}

\input{sections/experiments}

\input{sections/discussion}

\input{sections/related-work}

\input{sections/conclusion}

\input{sections/ack.tex}

\bibliography{references}
\bibliographystyle{IEEEtran}

\end{document}

%% file: header.tex
\usepackage{amsthm}
\usepackage{mdframed}
\usepackage{centernot}
\usepackage{comment}

\newcommand{\Aa}{\mathcal{A}}

\newcommand{\eE}{\mathbb{E}}
\newcommand{\Ff}{\mathcal{F}}

\newcommand{\Dd}{\mathcal{D}}

\newcommand\vd[2]{d_{i, p}}
\newcommand{\set}[1]{\left\{ #1 \right\}}

\newcommand{\seq}[1]{\langle #1 \rangle}

\newcommand{\R}{\mathbb R}

\newcommand{\Real}{\R}

\newcommand{\toolname}{\textsc{Dice}\xspace}
\usepackage{tikz}

\newtheorem{definition}{Definition}[section]
\newtheorem{example}{Example}

\newcommand{\qid}{\mathit{QID}}
\newcommand{\dnn}{\mathit{DNN}}

\definecolor{gold}{rgb}{0.99,0.78,0.07}
\usetikzlibrary{arrows,shapes,snakes,automata,backgrounds,positioning,decorations.pathmorphing,
	decorations.markings,calc}

\tikzstyle{dtreenode}=[draw=blue!10!gray,rounded rectangle, minimum size=5mm,fill=blue!10!white]
\tikzstyle{dtreeleaf}=[draw=black!60,minimum width=1cm,minimum height=0.4cm,rectangle,fill=blue!50!white]
\tikzset{every loop/.style={looseness=7}}
\tikzset{
	gluon/.style={decorate,draw=black,
		decoration={coil,amplitude=1pt, segment length=5pt}}
}
\tikzset{
	gluon1/.style={decorate,draw=black,
		decoration={coil,amplitude=3pt, segment length=3pt}}
}
\tikzset{
	gluonew/.style={decorate,draw=black,
		decoration={coil,amplitude=1pt, segment length=2pt}}
}

\tikzset{bicolor/.style args={#1 and #2 and #3}{
		path picture={
			\tikzset{rounded corners=0}
			\fill [#1] (path picture bounding box.south west)
			rectangle
			($(path picture  bounding box.north west)!#3!(path picture bounding
			box.north east)$);
			\fill [#2]
			($(path picture bounding box.south west)!#3!(path picture bounding
			box.south east)$)
			rectangle (path picture bounding box.north east);
}}}

\tikzset{tricolor/.style args={#1 and #2 and #3 and #4 and #5}{
		path picture={
			\tikzset{rounded corners=0}
			\fill [#1] (path picture bounding box.south west)
			rectangle
			($(path picture  bounding box.north west)!#4!(path picture bounding
			box.north east)$);
			\fill [#2]
			($(path picture bounding box.south west)!#4!(path picture bounding
			box.south east)$)
			rectangle
			($(path picture  bounding box.north west)!#5!(path picture bounding
			box.north east)$);
			\fill [#3]
			($(path picture bounding box.south west)!#5!(path picture bounding
			box.south east)$)
			rectangle (path picture bounding box.north east);
}}}

\usepackage[many]{tcolorbox}
\tcbuselibrary{listings,skins}

\lstdefinestyle{mystyle}{
  xleftmargin=0pt,
   basicstyle={\footnotesize\ttfamily},
   aboveskip=3mm,
   belowskip=3mm,
   keywordstyle=\bfseries,
   showstringspaces=false,
  escapechar=?,
  language=Java
}
\definecolor{code_indent}{HTML}{CCCCCC}

\newtcblisting{mylisting}[2][]{
  arc=0pt, outer arc=0pt,
  listing only,
  listing style=mystyle,
  title={\large #2},
  #1
}

 \definecolor{dkgreen}{rgb}{0,0.6,0}
 \definecolor{gray}{rgb}{0.5,0.5,0.5}
 \definecolor{mauve}{rgb}{0.58,0,0.82}

\definecolor{cadmiumgreen}{rgb}{0.0, 0.42, 0.24}
\definecolor{verde}{rgb}{0.25,0.5,0.35}
\definecolor{jpurple}{rgb}{0.5,0,0.35}
\definecolor{darkgreen}{rgb}{0.0, 0.2, 0.13}

\usepackage[ruled,vlined,linesnumbered,lined]{algorithm2e}
\usepackage{algpseudocode}

\usepackage{mathtools,xparse}

\usepackage{tabu}
\usepackage{multirow}

\usepackage{pifont}
\usepackage{enumerate}
\usepackage[shortlabels]{enumitem}

\usepackage{pgfplotstable}
\usepackage{pgfplots}

\usepackage[noframe]{showframe}
\usepackage{framed}

 {\endMakeFramed}
 \definecolor{shadecolor}{gray}{0.85}

\definecolor{bgblue}{RGB}{245,243,253}
\definecolor{ttblue}{RGB}{91,194,224}

\mdfdefinestyle{mystyle}{%
  rightline=true,
  innerleftmargin=10,
  innerrightmargin=10,
  outerlinewidth=3pt,
  topline=false,
  rightline=false,
  bottomline=false,
  skipabove=\topsep,
  skipbelow=false
}

\newtcolorbox{myboxi}[1][]{
  breakable,
  title=#1,
  colback=white,
  colbacktitle=white,
  coltitle=black,
  fonttitle=\bfseries,
  bottomrule=0pt,
  toprule=0pt,
  leftrule=3pt,
  rightrule=3pt,
  titlerule=0pt,
  arc=0pt,
  outer arc=0pt,
  colframe=black!50,
}

\newtcolorbox{myboxii}[1][style=mystyle]{
  breakable,
  freelance,
  colback=white,
  colbacktitle=white,
  coltitle=black,
  fonttitle=\bfseries,
  bottomrule=0pt,
  boxrule=0pt,
  colframe=white,
  after skip=0pt,
  overlay unbroken and first={
    \draw[white!75!black,line width=3pt]
    ([yshift=-9pt]frame.north west) --
    ([yshift=9pt]frame.south west);
  },
}

%% file: sections/introduction.tex
\section{Introduction}
\label{sec:introduction}

AI-assisted software solutions---increasingly implemented as deep neural networks~\cite{GoodBengCour16} (DNNs)---have made substantial inroads into critical software infrastructure where they routinely assist in socio-economic and legal-critical decision making~\cite{Ranchords2021AutomatedGF}. 
Instances of such AI-assisted software include software deciding on recidivism, software predicting benefit eligibility, and software deciding whether to audit a given taxpayer.
The DNN-based software development, driven by the \emph{principle of information bottleneck}~\cite{tishby2000information}, involves a delicate balancing act between over-fitting and detecting useful, parsimonious patterns.
It is, therefore, not a surprise that such solutions often encode and amplify pre-existing biases in the training data. What's worse, improper training may even introduce biases not present in the training data or irrelevant to the decision making.
The resulting fairness defects may not only disadvantage protected groups~\cite{compas-article,elyounes2020computer,DBLP:journals/pacmhci/EscherB20,slemrod2022group,dorothyIRS}, but may stand to violate statutory requirements~\cite{thomas2022automating}.
\begin{myboxii}
\emph{This paper presents \toolname{},
an information-theoretic testing and debugging framework for fairness defects in deep neural networks.}
\end{myboxii}

\vspace{0.5em} \noindent\textbf{Quantifying Fairness.} Concentrated efforts from the software engineering and the machine learning communities have produced a number of successful \emph{fairness testing} frameworks~\cite{angell2018themis,galhotra2017fairness,aggarwal2019black,sharma2020automatic}. 
These frameworks characterize various notions of fairness---such as group fairness~\cite{hardt2016equality} (decision outcome for various protected groups must be similar) and individual fairness~\cite{dwork2012fairness} (individuals differing only on protected attributes must receive similar outcome)---and employ search-based testing to discover fairness defects. 
While a binary classification of fairness is helpful in discovering defects, developers may require further insights into the nature of these defects to decide on the potential ``bug fix''.
Are some defects more severe than others? Whether these defects stem from biases present in the training data, or they are artifacts of an inadequate training? Is it possible to find an alternative explanation of the training data that does not use protected information?

\emph{Individual discrimination} is a  well-studied~\cite{10.1145/3510003.3510137,galhotra2017fairness,zhang2020white,9793943} causal notion of fairness that defines a function being discriminant towards an individual (input) if there exists another individual (potentially counterfactual), differing only in the protected features, receives a more favorable outcome.
We present a quantitative generalization of this notion as the \emph{quantitative individual discrimination} (QID). We define QID as the amount of protected information---characterized by entropy metrics such as Shannon entropy and min entropy---used in deriving an outcome. 
Observe that a zero value for the QID measure implies the absence of the individual discrimination.
The QID measure allows us to order various discriminating inputs in terms of their severity, as in an application that is not supposed to base its decisions on protected information, inputs with higher dependence indicate a more severe violation.
Our first \emph{research question} ({\bfseries RQ1}) concerns the usefulness of QID measure in finding inputs with different severity. 

\vspace{0.5em} \noindent\textbf{Search-Based Testing.} 
Search-based software testing  provide scalable optimization algorithms to automate discovery of software bugs. 
In the context of fairness defects, the search of such bugs
involves finding twin inputs exhibiting discriminatory instances.
The state-of-the-art algorithms for fairness testing~\cite{udeshi2018automated,zhang2020white,9793943} explore the input space governed by a binarized feedback, resulting in a discontinuous search domain. On the other hand, QID-based search algorithms can benefit a smooth (quantitative) feedback during the optimization, resulting in a more guided search.
Our next research question ({\bfseries RQ2}) is to investigate whether this theoretical promise materializes in practice in terms of discovering richer discriminating instances than using classic notions of discrimination.

\vspace{0.5em} \noindent\textbf{Causal Explanations.}
While the discriminating instances (ordered by their severity) provide a clear evidence of fairness defects in the DNN, it is unclear whether these defects are inherent in the training data, or whether they are artifacts of the training process. 
Inspired by the notion of ``the average causal effects''~\cite{glymour2016causal} and \textsc{Audee} framework~\cite{guo2020audee} for bug localization in deep learning models, we develop a layer and neuron localization framework for fairness defects. If the cause of the defects is found to be at the input layer, it is indicative of discrimination existing in the training data.  
On the other hand, if we localize the cause of the defect to some internal layer, we wish to further prod the DNN to extract quantitative information about neurons and their counterfactual parameters that can mitigate the defect while maintaining the accuracy. 
This debugging activity informed our next research question ({\bfseries RQ3}): 
is it possible to identify a subset of neurons and their causal effects on QID to guide a mitigation without affecting accuracy? 

\vspace{0.5em}\noindent \textbf{Experiments.} \toolname{} implements a search algorithm (Algorithm~\ref{alg:overall-search}) to discover inputs that maximize QID and a causal debugging algorithm (Algorithm~\ref{alg:overall-debugging}) to localize layers and neurons that causally affect the amounts of QID. 
Using $10$ socio-critical DNNs from the literature of algorithmic fairness, we show that \toolname{} finds inputs
that can use significant amounts of protected information in the decision making; outperforms three
state-of-the-art techniques~\cite{udeshi2018automated,zhang2020white,9793943} in generating discriminatory instances;
and localizes neurons that guides a simple mitigation strategy to reduce QID down to $15\%$ of reported initial QID with at most $5\%$ loss of accuracy. 
The key contributions of this paper are:
\begin{enumerate}
    \item {\bf Quantitative Individual Discrimination.} We introduce an information-theoretic characterization of discrimination, dubbed quantitative individual discrimination (QID), based on Shannon entropy and Min entropy. 
    \item {\bf Search-based Testing.} We present a search-based algorithm to discover circumstances under which the DNNs exhibit severe discrimination.
    \item {\bf Causal Debugging.} We develop a causal fairness debugging based on the language of interventions to localize the root cause of the fairness defects.
    \item {\bf Experimental Evaluation.} Extensive experiments over different datasets and DNN models that show feasibility, usefulness, and scalability (viz-a-viz state-of-the-art). Our framework can handle multiple protected attributes and can easily be adapted for regression tasks. 
\end{enumerate}

%% file: sections/background.tex
\section{Preliminaries}
\label{sec:background}

\noindent \textbf{Fairness Terminology.}
\label{subsec:Fairness-Terminology}
We consider decision support systems as \textit{binary classifiers} where a prediction label is \textit{favorable} if it gives a desirable outcome to an input (individual). These favorable predictions may include higher income
estimations for loan, low risk of re-offending in parole assessments,
and high risk of failing a class. 
Each dataset consists of a number of \textit{attributes} (such as income,
experiences, prior arrests, sex, and race) and a set
of \textit{instances} that describe the value of attributes for each individual.
According to ethical and legal requirements, data-driven software should not \textit{discriminate}
on the basis of an individual's \emph{protected attributes} such as sex, race, age, disability,
colour, creed, national origin, religion, genetic information, marital status, and sexual orientation.

There are several well-established fairness definitions. \textit{Group fairness} requires that the statistics of ML outcomes for different \emph{protected groups} to be similar~\cite{hardt2016equality} using metrics
such as \textit{equal opportunity difference} (EOD), which is the difference between
the true positive rates (TPR) of two protected groups. Fairness through unawareness (FTU)~\cite{dwork2012fairness} requires removing protected attributes during training. 
However, FTU may provide inadequate support since protected attributes can influence the prediction via a non-protected collider attribute (e.g., race and ZIP code).
Fairness through awareness (FTA)~\cite{dwork2012fairness} is
an \textit{individual fairness} notion that requires that two \textit{individuals} that deemed similar (based on their non-protected attributes) are treated similarly. \textit{Our approach is geared toward individual fairness.}

\vspace{0.5em}\noindent \textbf{Individual Discrimination.}
Causal discrimination, first studied in \textsc{Themis}~\cite{angell2018themis},
measures the difference between two subgroups via \textit{counterfactual} queries. 
It samples individuals with the protected attributes set to $A$ and compares the outcome to a counterfactual scenario where the protected attributes is set to $B$.
Individual discrimination (ID) is a prevalent notion that adapts counterfactual queries to find an individual such that their counterfactual with a different protected attributes receives more favorable outcome. 
This fairness notion is used by
the state-of-the-art fairness testing to generate fairness defects~\cite{agarwal2018automated,udeshi2018automated,zhang2020white,9793943} and closely related to situation testing notion~\cite{10.1145/3468264.3468537}.
While standard group fairness metrics (e.g., AOD/EOD) are already quantitative, the quantitative measures do not exist for individual fairness. We propose to adapt information theoretic tools to provide quantitative measures for individual fairness.

\vspace{0.5em}\noindent \textbf{Information-Theoretic Concepts.}
\label{subsec:Quantitative-Notion}
The notion of R\'enyi entropy~\cite{renyi1961measures}, $H_\alpha(X)$ quantifies the uncertainty (randomness) of a system responding to inputs $X$. 
In particular, Shannon entropy ($\alpha{=}1$) and min-entropy ($\alpha{=}\infty$) are two important subclasses of R\'enyi entropy. Shannon entropy ($H_1$) measures the expected amounts of uncertainty over finitely many events whereas min entropy ($H_{\infty}$) measures the uncertainty over single maximum likelihood event. 

Consider a deterministic system (like pre-trained DNN) with a finite set of responses and assume that the input $X$ is distributed uniformly. Thus, the system induces an equivalence relation over the input set $X$ such that two inputs are equivalent if their system outputs are approximately close, i.e. $x {\sim} x'$ iff $\dnn(x)\approx_{\epsilon}\dnn(x')$. Let $X_o$ denote the equivalence class of $X$ with output $o$. Then, the remaining uncertainty after observing the output of DNN over $X$ can be written as:
\[
H_1(X|O)=\sum_{O=o}\frac{|X_o|}{|X|}.\log_{2}(|X_o|)~~~~~~~~(\text{Shannon entropy})
\]
where $|X|$ is the cardinality of $X$ and $|X_o|$ is the size of equivalence class of output $o$. Similarly, the min-entropy is given as
\[ 
H_{\infty}(X|O)=\log_{2}(\frac{|X|}{|O|})~~~~~~~~~~~~~~~~~~~~~~~~~(\text{min-entropy})
\]
where $|O|$ is the number of equivalence classes over $X$~\cite{10.1007/978-3-642-00596-1_21,backes2009automatic,10.1145/3460319.3464817}. Given that the initial entropy is equal to $\log_2(|X|)$ for both entropies, the amount of information from $X$ used by the system to make decisions are 
\begin{eqnarray*}
I_1(X;O) &=& \log_2(|X|) - H_1(X|O), \text { and }\\ I_{\infty}(X;O) &=& \log_{2}(|O|),
\end{eqnarray*}
under Shannon- and min- entropies with $I_{1} \leq I_{\infty}$.

\vspace{0.5em}\noindent \textbf{Quantitative Notion and Fairness.}
\label{subsec:Quantitative-Notion-Fairness}
Our approach differs from these state-of-the-art techniques~\cite{10.1145/3338906.3338937,agarwal2018automated,udeshi2018automated,zhang2020white,10.1145/3460319.3464820,9793943}
in that it extends the individual discrimination notion with quantitative information flow 
that enables us to measure the amount of discrimination in the ML-based software systems.
Given non-protected attributes and ML outcomes, 
the \textit{Shannon entropy} measures the expected amount of individual discrimination over all possible responses varying protected classes, whereas, \textit{min entropy} measures the amount over a single response from the maximum likelihood class.

\begin{example}
Consider a dataset with $16$ different
protected values---sex(2), race(2), and age(4)---distributed uniformly,
and suppose that we have $4$ individuals in the system. We perturb the protected attributes of these individuals to
generate $16$ counterfactuals and run them through the DNN to get their prediction scores.
Suppose that the outputs have all $16$ outputs in the same class for the first individual (absolutely fair) and
have $16$ classes of size one for the second individual (absolutely discriminatory).
For the third individual, let the outputs be in $4$ classes with $\set{4,4,4,4}$
elements in each class (e.g., there is one output class per each age group).
For the fourth individual, let us consider outputs to be in $5$ classes with $\set{8,4,2,1,1}$
elements in each class (e.g., \texttt{if} race=1 \texttt{then} the output is class $1$; else,
\texttt{if} sex=1 \texttt{then} the output is $2$; else
\texttt{if} age=\{1,2\} \texttt{then} the output is $3$; else
there is one output class for each age=\{3,4\}).
\end{example}

We work with the following notions of discrimination.
\begin{itemize}
    \item 
\noindent \textit{Individual Discrimination Notion}. The individual discrimination used by the
state-of-the-art techniques can only
distinguish between the first individual and the rest, but they cannot distinguish
among individuals two to four. In fact, these techniques generate
tens of thousands of individual discriminatory instances in a short amount of time~\cite{udeshi2018automated,zhang2020white,9793943}.
However, they fail to prioritize test cases for mitigation and
cannot characterize the amounts of discrimination (i.e., their severity).

\item \noindent \textit{Shannon Entropy}. Using the Shannon entropy, we have the initial
fairness to be $4.0$ bits, a maximum possible discrimination. The remaining fairness
of DNN are $4.0$, $0.0$, $2.0$ and $2.125$, for the first to fourth individuals, respectively.
The discrimination is the difference between the initial and remaining fairness,
which are $0.0$, $4.0$, $2.0$ and $1.875$ for the first to fourth individuals, respectively.
It is important to note that beyond the two extreme cases, Shannon entropy deems perturbations
to the third individual (rather than the fourth) create a higher amount of discrimination. 

\item \noindent \textit{Min Entropy}. The initial fairness via min entropy is also $4$ bits.
The conditional min entropy is $\log \frac{16}{1}=4.0$, $\log \frac{16}{16}=0.0$,
$\log \frac{16}{4}=2.0$, and $\log \frac{16}{5}=1.7$, for the four individuals, respectively.
The amounts of discrimination thus are $0.0$, $4.0$, $\log 4=2.0$ and $\log 5=2.3$, respectively. Beyond the two extreme cases where both entropies agree,
the min entropy deems perturbations to the fourth individual create a higher amount of discrimination. This is intuitive since the discrimination on the fourth case is more subtle, complex, and significant. Therefore, the ML software developers might prioritize those cases characterized by the min entropy.
\end{itemize}

%% file: sections/overview.tex
\section{Overview}
\label{sec:overview}
\begin{figure*}
    \centering
    \includegraphics[width=\textwidth]{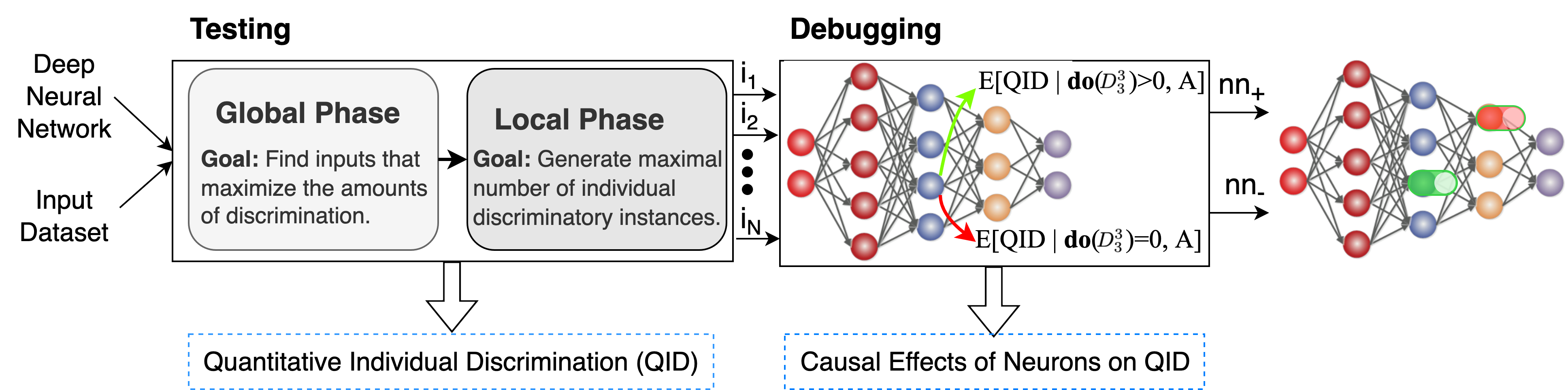}
    \caption{Workflow of \toolname. Given a DNN and relevant input dataset, \toolname quantifies QID discrimination via testing and applies causal debugging to localize and mitigate QID discrimination.}
    \label{fig:overview-QID}
\end{figure*}

\noindent \textbf{\toolname in Nutshell.} Figure~\ref{fig:overview-QID} shows an overview of our framework \toolname.
It consists of two components: (1) an automatic test-generation mechanism based on
search algorithms and (2) a debugging approach that localizes the neurons with significant impacts on fairness
using a causal algorithm. First, \toolname searches through the space of input dataset
to find circumstances on the non-protected attributes under which the DNN-under-test shows a significant dependency on
the protected attributes in making decisions. 
In doing so, it works in global and local phases. In the
global phase, the search explores the input space to increase the amount of discrimination in each step
of search. On the other hand, in the local phase it exploits the promising seeds from the global phase to generate as many discriminatory instances as possible. 

The key elements of search is a threshold-based clustering algorithm used for computing both gradients and objective functions that provide a smooth feedback.
The search characterizes the quantitative individual discrimination (QID)
and returns a set of interesting inputs. Second, \toolname uses those inputs to localize neurons with
the largest causal effects on the amounts of discrimination. In doing so, it intervenes~\cite{pearl2009causality} over a set of
suspicious neurons.
For every neuron, our debugging approach forces the neurons to be active ($n{>}0$) and non-active ($n{=}0$) over the test cases as far as the functional accuracy of DNN remains in a valid range. Then, it computes the difference between the amounts of QID in these two cases to characterize the causal effects of the neuron on fairness.

\toolname{} reports top $k$ neurons that both have positive impacts (i.e., their activation
reduces the amounts of discrimination) and negative impacts (i.e, their activation
increases the amounts of discrimination). 
A potential mitigation strategy is to intervene
to keep a small set of neurons activated (for the positive neurons) or deactivated (for the negative neurons). 

\vspace{0.5em}
\noindent \textbf{Test Cases.} 
Consider the \emph{adult census income}~\cite{Dua:2019} dataset with a pre-trained model with $6$ layers~\cite{zhang2020white} to overview \toolname in practice. We ran \toolname for $1$ hours and obtain $230,593$ test cases. It discovered $36$ clusters
from the initial of $14$ clusters, and the amounts of QID are $4.05$ and $2.64$ bits
for min entropy and Shannon entropy, respectively, out of a total of $5.3$ bits of information from the protected attributes. Considering to order the test cases, we have $6$ test case with maximum QID discrimination of $5.3$ bits. In addition, we have $29$ and $112$ test cases cases with $5.2$ and $5.1$ bits of QID discrimination. The reported numbers are averaged (and rounded) for $10$ runs.

\vspace{0.5em}
\noindent \textbf{Localization and Mitigation.} \toolname uses the generated test cases to localize layers and neurons with a significant causal contribution
to the discrimination. For the census dataset, it identifies the second layer as the layer with largest sensitivity to protected attributes.
Among the neurons in this layer,
\toolname found that $15$th neuron has the largest negative influence on fairness (the discrimination decreased by $19.6\%$ when it is deactivated) and $19$the neuron has the largest positive influence on fairness (the discrimination decreased by $17.6\%$ when it is activated).
Following this localization, a simple mitigation strategy of activating or deactivating these neuron reduces the amounts of QID discrimination by
$20\%$ with $3\%$ accuracy loss.

\vspace{0.5em}
\noindent \textbf{Comparison to the State-of-the-art.} We compare \toolname to the
state-of-the-art techniques in terms of generating individual discrimination (ID) (rather than the quantitative notion)
per each protected attribute. Our goal is to evaluate whether the clustering-based search is effective in generating discriminatory instances.
We run \toolname and baseline for $15$ minutes, and report average of results over $10$ runs. 
The baseline includes \textsc{Aequitas}~\cite{udeshi2018automated}, \textsc{ADF}~\cite{zhang2020white}, 
and \textsc{NeuronFair}~\cite{9793943}. Considering sex as the protected attribute in the census dataset, \toolname generated $79.0k$ instances whereas \textsc{Aequitas}, \textsc{ADF}, and \textsc{NeuronFair} generated $10.4k$, $18.2k$, and $21.6k$ discriminatory instances, respectively. 
Overall, \toolname generate more ID instances in all cases with more success rates. However, \toolname is slower in finding the first ID
instance in order of a few seconds (in average), since our approach does not generate ID instances in global phase. When considering the time to the first 1,000 instances, \toolname has significantly outperformed the state-of-the-art. We conjecture that the improvements are due to smooth search space via quantitative feedback.

%% file: sections/definition.tex
\section{Problem Statement}
\label{sec:definition}
We consider DNN-based classifiers with the set of input variables $A$ partitioned into protected set of variables $Z$ (such as race, sex, and age) and non-protected variables $X$ (such as profession, income, and education). 
We further assume the output to consist of $t$ prediction classes.

\begin{definition}[DNN: Semantics]
  A deep neural network (DNN) encodes a function $\Dd: X \times Z \to [0,1]^t$ where $X = X_1 \times X_2 \cdots \times X_n$ is the set of non-protected input variables, $Z = Z_1 \times Z_2 \cdots \times Z_r$ is the set of protected input variables, and 
  the output is $t$-dimensional probabilistic vector corresponding to $t$ prediction classes. The predicted label $\Dd_{\tt \ell}(x, z)$ of an input pair $(x, z)$ is the index of the maximum score, i.e. $\Dd_{\tt \ell}(x, z) = \max_{i} \Dd_{\tt \ell}(x, z)(i)$. 
  We assume that the set of protect input variables are finite domain, and we let $m$ be the cardinality of the set of protected variables $Z$.
\end{definition}

\begin{definition}[DNN: Syntax]
A DNN $\Dd$ is parameterized by the input dimension $n{+}r$, the output dimension $t$, the depth of hidden layers $N$,  and the weights of its hidden layers $W_1, W_2, \ldots, W_N$. Our goal is to test and debug a pre-trained neural network with known parameters and weights. 
Let $D_{i}$ be the output of layer $i$ that implements an affine mapping from the output of previous layer $D_{i-1}$ and its weights $W_{i-1}$ for $1 \leq i \leq N$ followed by
\begin{enumerate}
    \item 
     a fixed non-linear activation unit (e.g., ReLU defined as $D_{i-1} \mapsto \max \set{W_{i-1}.D_{i-1}, 0}$) for $1 \leq i < N$, or
     \item
     a SoftMax function that maps scores to probabilities of each class for $i=N$. 
\end{enumerate} Let $D_i^{j}$ be the output of neuron $j$ at layer $i$.
\end{definition}

\vspace{0.5em}
\noindent \textbf{Individual Discrimination.} We say a DNN $\Dd$ is biased based on causal discrimination notion~\cite{10.1145/3510003.3510137,galhotra2017fairness,zhang2020white,9793943} if 
\[
\exists z_1, z_2 \in Z, x \in X~s.t.~ \Dd_\ell(x, z_1) \neq \Dd_\ell(x, z_2),
\]
for $z_1 \neq z_2$ of protected inputs. Intuitively, the idea is to find an individual such that their counterfactual with different protected attributes such as race receives a different outcome.

\vspace{0.5em}
\noindent \textbf{Quantitative Individual Discrimination.} In the setting of fairness testing,
it is often desirable to quantify the \textit{amounts} of bias for individuals. We define the notion quantitative individual discrimination (QID) based on the equivalence classes induced from the output of DNN over protected attributes. Formally, $\qid(Z,X=x) = \seq{Z_1,\ldots,Z_k}$ that is the quotient space of $Z$ characterized by the DNN outputs under an individual with non-protected value $x$. Using this notion, we say a pair of protected values $z,z'$ are in the same equivalence class $i$ (i.e., $z,z' \in Z_i$) if and only if $\Dd(z,x)\approx\Dd(z',x)$.

Given that $Z$ is uniformly distributed
and $\Dd$ is a deterministic function,
we can quantify the QID notion for an individual $(z,x)$ according to the Shannon
and min entropy, respectively:
\[
Q_1(Z,x) {=}  \log_2(m){-}\sum_{i=1}^{k} \frac{|\qid_i(Z,x)|}{m}.\log_2(|\qid_i(Z,x)|)
\]
\[
Q_{\infty}(Z,x) = \log_2(m) - \log_2(\frac{m}{k}) = \log_2(k).
\]
where $m$ is the cardinality of $Z$,
$|\qid_i(Z,x)|$ is the size of equivalence class $i$, and 
$k$ is the number of equivalence classes.

\vspace{0.5em}
\noindent \textbf{Debugging/Mitigating DNN for QID.} After characterizing the amounts of discrimination via $\qid$, our next step is to localize a set of layers and neurons that \textit{causally} effect the output of DNN to have $k$ equivalence classes. 

Causal logic~\cite{pearl2009causality} provides a firm foundation to reason about the causal relationships between variables. We consider a structural causal model (SCM) with exogenous variables $U$ over the unobserved
input factors, endogenous variables $V$ over $(X,Z,D_i^j)$; and the set of functions $\Ff$ over the set $V$ using the DNN function $\Dd$ and exogenous variables $U$. Using the SCM, we aim to estimate the average causal effect (ACE)~\cite{pearl2009causality} of neuron $D_i^j$ on the QID. 

A primary tool for performing such computation is called \texttt{do} logic~\cite{glymour2016causal}. We write \texttt{do}($i,j,y$) to indicate that the output of neuron $j$ at layer $i$ is intervened to stay $y$. In doing so, we remove the incoming edges to the neuron and force the output of neuron to take a pre-defined value $y$, but we are not required to control back-door variables due to the feed-forward structure of DNN. Then, the ACE of neuron $D_i^j$ on the quantitative individual discrimination with min entropy can be written as $E[Q_{\infty}|$\texttt{do}($i,j,y$),$k,l]$,
which is the expected QID after intervening on the neuron given that the non-intervened DNN characterized $k$ classes with an accuracy of $l$. Our goal is to find neurons with the largest causal effects on the QIDs, requiring that such interventions are faithful to the functionality of DNN. 

\begin{definition}[Quantitative Fairness Testing and Debugging]\label{def:problem-2}
Given a deep neural network model $\Dd$ trained over a dataset $A$ with
protected ($Z \subset A$) and non-protected ($X \subset A$) attributes;
the search problem is to find a single non-protected value $x \in X$
such that the quantitative individual discrimination (QID), for a chosen measure $Q_1$ or $Q_{\infty}$,
is maximized over the $m$ protected values $\Sigma=\{z_1,\ldots,z_m\}$.
Given the inputs $(\Sigma,x)$ characterizing the maximum QID,
our debugging problem is to find a minimal subset of layers
$l \subset \{1,\ldots,N\}$ and neurons $D_{l}^{(J)}$ for $J \subseteq |W_{l}|$
such that the average causal effects of $D_l^{J}$ on the QID
are maximum. 
\end{definition}

%% file: sections/approach.tex
\section{Approach}
\label{sec:approach}

\noindent \textbf{Characterizing Quantitative Individual Discrimination.}
Given a DNN $\Dd$ over a dataset $A$, our goal is to characterize the worst-case
QID over all possible individuals. Since min entropy characterizes the amounts of discrimination
from one prediction with $Q_{\infty}(Z,x)\geq Q_1(Z,x)$, it is a useful notion to prioritize
the test cases. Therefore, we focus on $Q_{\infty}$ and propose the following objective function:
\begin{center}
$\max_{x \in X}~2^{Q_\infty(Z,x)} + (1 - exp(-0.1*\delta))$
\end{center}
where $2^{Q_\infty(Z,x)}=k$ and $\delta$ is the maximum distance between equivalence classes, normalized with the exponential function to remain between $0$ and $1$.
The term is used to break ties when two instances characterize the same number of classes, by preferring one with the highest distance. 
Overall, the goal is to find a single value of non-protected attribute $x$ such that the neural network model $\Dd$ predicts many distinguishable
classes of outcomes when $x$ is paired with $m$ protected values.
However, finding those inputs requires an exhaustive search in the exponential set of subsets of input space,
and hence is clearly intractable. We propose a gradient-guided search algorithm that aims to search the space of input variables (attributes) to
maximize the number of equivalence classes and generate as many discrimination instances as possible. 

\vspace{0.5em}
\noindent \textbf{Search Approach.} 
Our search strategy
consists of global and local phases as in some of the prior work~\cite{zhang2020white,10.1145/3460319.3464820,9793943}. 
The goal of the global phase is to find the maximum quantitative individual discrimination via gradient-guided clustering. The local phase uses the promising instances to generate a maximum number of discriminatory instances (ID). 

\vspace{0.25em}\noindent \textit{Global Phase.} 
Given a current instance $x$, the global stage first uses $m$ different values from the space of protected attributes, while keeping the values of non-protected attributes the same.
Then, it receives $m$ prediction scores from the DNN and partitions them into $k$ classes. We adapt a constrained-based clustering with $\epsilon$ where two elements cannot be in the same cluster if their scores differ more than $\epsilon$. Now, the critical step is to perturb the current instance over a subset of non-protected attributes with a direction
that will likely increase the number of clusters induced from the perturbed instance in the next step of global search.

In doing so, we first compute the gradients of DNN loss function for a pair of instances (say $a,a'$) in the cluster with the maximum
elements. The intuition is that we are more likely to split the largest cluster into $2$ or more sub-clusters and increase
the number of partitions in the next step. For the pair of samples, we use the non-protected attributes that
have the same direction of gradients $d$ since it shows the high sensitivity of loss function with respect to
small changes on those common features of the pair.
If we were to use gradients of opposite directions, we will neutralize
the effects of gradients since we only perturb one instance over the non-protected attributes.
Finally, we perturb the current sample $x$ to generate $x'$ using the direction $d$ and step size $s_g$.

\vspace{0.25em}\noindent \textit{Local Phase.}
Once we detect an instance with more than $2$ clusters, we enter a local phase where the goal
is to generate as many discriminatory instances (ID) as possible. In our quantitative approach,
we say that an unfavorable decision for an individual $x$ is discriminatory if there is a counterfactual
individual $x'$ that received a favorable outcome. Similar to the state-of-the-art~\cite{udeshi2018automated,zhang2020white,9793943}, we use non-linear
optimizer that takes an initial instance $x$, a step function to generate the next instance around the neighborhood
of the current instance, and an objective function that quantifies the discrimination of the current instance.
Since our approach uses a continuous objective 
based on the characteristics of clusters, it enables us to guide the local search to generate discriminatory instances.

\input{sections/Algorithm1}

\vspace{0.5em}
\noindent \textbf{Search Procedure.} 
Algorithm~\ref{alg:overall-search} sketches our search algorithm to quantify the amounts of bias. We first use the clustering (KMeans algorithm)
to partition the data points into $p$ groups (line 1). 
Next, we run the algorithm until the time-out $T$ reaches where in each iteration we seed a sample randomly from one of the partitions $p$ (line 2). 
Then, we proceed into global and local phases of
search. 

In the global phase, we first use the seed instance $x$ and perturb the protected attributes $P$ to generate $m$
possible instances $X_m$ with different protected values and the same non-protected ones. Then, we run $X_m$
through the DNN model to get the probability scores $S_m$ (line 5). Then, we cluster the scores $S_m$ into
$k$ groups using the tolerance $\epsilon$ (line 6). Afterward, we compute the gradients over two random instances
(from the cluster with the largest size) and use a subset of non-protected features that have the largest number of agreements
on the gradient directions. Then, we use those features and their directions to 
perturb the inputs $x$ and generate the next instance $x'$ (line 7-10).
Then, we enter the local phase if the number of clusters or distance between are increased (line 11). 

In the local phase, we use the general-purpose optimizer, known as \texttt{LBFGS}~\cite{liu1989limited},
which takes an initial seed $x$, an objective function, a step function, and the maximum number of local iterations $N_l$;
it returns the generated instances during the optimization (Line 12-19).
In the objective function shown with \texttt{eval\_f} (Line 12-17),
we generate $m$ instances with the same non-protected values but different protected ones (Line 13).
We generate prediction scores for those instances and cluster them with tolerance parameter $\epsilon$ (line 14).
Then, we compute the difference between the indices of two clusters with the smallest and largest scores (line 15).
Finally, we record the generated sample and return the difference as the evaluation of optimizer at the current sample
(line 16-17). The step function is shown with \texttt{perturb\_local} (Line 18) where it guides the
optimizer to take one step in the input space. Our step function uses a random sample
from a different cluster compared to the current sample. Then, it computes the
normalized sum of gradients and perturbs it using the smallest gradients
to remain in the neighborhood of the current sample.

\vspace{0.5em}
\noindent \textbf{Debugging Approach.} Since it is computationally difficult to intervene over all possible neurons in a DNN, we first adapt a layer-localization technique from the literature of DL framework debugging~\cite{pham2019cradle,guo2020audee} where we detect a layer with the largest sensitivity to the protected attributes. Let $D_i(z,x)$ be the output of layer $i$ over protected value $z$ and non-protected value $x$.
Let $\Delta_i(x): \Real^{|D_i|} \times \Real^{|D_i|} \to \Real$
be the distance between the outputs of DNN at layer $i$ as triggered by
$m$ different protected values and the same non-protected value $x$,
and let $\delta_i$ be the $\max_{x}\Delta_i(x)$. 
The rate of changes in the sensitivity of layer $i$ (w.r.t protected attributes) is
\[
\rho_i = \frac{\delta_i - \max_j\delta_j}{\max_j\delta_j+\epsilon},~with~0\leq~j<~i
\]
where $\delta_0=0.0$ and $\epsilon=10^{-7}$ (to avoid  division-by-zero~\cite{pham2019cradle,guo2020audee}). Let $l = \argmax_i \rho_i$ be the layer index with the maximum rate of changes. Our next step is to localize neurons in the layer $l$ that have significant positive or negative effects on fairness. Let $V_l^j$ be the set of possible values for neuron $j$ at later $l$ (recorded during the layer localization). 
We are interested in computing the average causal effects when the neuron $D_l^j$ is activated vs. deactivated, noting that such interventions might affect the functionality of DNN. Therefore, among a set of intervention values,
we choose one activated value $v_1\in V_l^j > 0$ and one deactivated value $v_2\in V_l^j\approx0$, considering the functional accuracy of DNN within $\epsilon$ of original accuracy $A$.
Therefore, we define average causal difference (ACD) for a neuron $D_l^j$ as:
\[
\eE[Q_{\infty} \mid \texttt{do}(l,j,v_1>0),k,A]{-}\eE[Q_{\infty} \mid \texttt{do}(l,j,v_2\approx0),k,A],
\]
where \texttt{do} notation is used to force the output of neuron $j$ at layer $l$ to a fix value $v$.
We then return the neuron indices with the largest positive (aggravating discrimination) and smallest negative (mitigating discrimination).
Let $\hat{i}$ and $\hat{j}$ be the layer and neuron with the largest positive $ACD$.
One simple mitigation strategy is thus to
deactivate the neuron $D_{l}^{\hat{j}}$, expecting to reduce QID by $ACD/k$ percentage.
Similarly, activating the neuron with the smallest negative $ACD$ is expected
to reduce QID by $ACD/k$ percentage.

\input{sections/Algorithm2}

\vspace{0.5em}
\noindent \textbf{Debugging Procedure.} Algorithm~\ref{alg:overall-debugging} shows the debugging aspect of \toolname.
Given a set of test cases from the search algorithm, we first use a notion of distance (e.g, $\Delta=L_1$) to compute
the difference between any pair of protected values $z,z'$ w.r.t the outputs of layer $l \in \{1,\ldots,N\}$ (line 1).
Then, we compute the rate of changes (line 2-3) and return a layer $l$ with the largest change (line 4). We compute various
statistics on the output of every neuron $i$ at layer $l$ such as the minimum, maximum, average, average{$\pm$}std. dev, average{$\pm2*$}std. dev, etc (line 5).
Among those values, we take the smallest and largest values such that the intervention on the neuron $i$ at layer $l$ has the minimal impacts on the accuracy of DNN (line 6).
Finally, we compute the average causal difference (line 7) and return the indices of layer, neurons with large negative influence, and neurons with large positive influence.

%% file: sections/Algorithm1.tex
\begin{algorithm}[t!]
{
	\DontPrintSemicolon
	\KwIn{Dataset $A$, deep learning model $\Dd$,
        the loss function for the DNN $J$,
	protected attributes $P$, non-protected attributes $NP$, 
        the number of partitions over the dataset $p$,
        the step size in global perturbation $s_g$,
        the step size in local perturbation $s_l$,
        the maximum number of global iterations $N_g$,
        the maximum number of local iterations $N_l$,
	the tolerance $\epsilon$, and time-out $T$.
	}
	\KwOut{Num. Clusters and (local+Global) Test Cases.}

	$A'$, $cur$ $\gets$ \texttt{KMeans}($A$, $p$), \texttt{time}()

	\While{\texttt{time}() - $cur$ $<$ $T$}{
        $x$, $i$, $k$, $\delta$ $\gets$ \texttt{pick}($A^{'}$), $0$, $1$, $0.0$

        \While{$i<N_g$}{
            $I_m, S_m$ $\gets$ \texttt{Generate\_Predict}($a$, $P$)\\
            $X_k$, $\delta'$ $\gets$ \texttt{Clust}($S_m$, $\epsilon$)\\
            $a, a'$ $\gets$ \texttt{Choose\_Pair\_Max}($X_k$)\\
            $Gs$ $\gets$ $({\nabla}J(a), {\nabla}J(a'))$\\
            $d$ $\gets$ \texttt{choose\_common\_direct}($Gs$, $NP$)\\
            $x'$ $\gets$ \texttt{perturb}($x, d, s_g$)\\
                \If{($|X_k|$ $>$ $k$) or ($|X_k|=k$ and $\delta' > \delta$)}{
                    \texttt{eval\_f}($x$) $\{$ \\
                    \hspace{0.5em}$I'_m, S'_m$ $\gets$ \texttt{Generate\_Predict}($x$, $P$)\\
                    \hspace{0.5em}$X_{k'}$ $\gets$ \texttt{Clust}($S'_m$, $\epsilon$)\\
                    \hspace{0.5em}$\Delta$ $\gets$ $arg.\max$($X_{k'}$)${-}arg.\min$($X_{k'}$)\\
                    \hspace{0.5em}$local\_inps$.\texttt{add}($x$)\\
                    \hspace{0.5em}\textit{Return} $-\Delta$\} \\        
                    \texttt{step\_f} $\gets$ $\lambda_x$ \texttt{perturb\_local}($x, s_l$) \\
                    \texttt{LBFGS}($x$, \texttt{eval\_f}, \texttt{step\_f}, $N_l$)  \\ 
                }
            $global\_inps$.\texttt{add}($a$)\\
            $k$, $x$ $\gets$ \texttt{max}($k, |X_k|$), $x'$\\ 
        }
    }    
    \Return $k$, $I=global\_inps \cup local\_inps$

\caption{\textsc{\toolname (Search)}}
\label{alg:overall-search}
}
\end{algorithm}

%% file: sections/Algorithm2.tex
\begin{algorithm}[t!]
{
	\DontPrintSemicolon
	\KwIn{Dataset $A=(A_X,A_Z)$, $\Dd$ with accuracy $\Aa_{\Dd}$,
        Test cases $I$,
        the distance function $\Delta$, the tolerance of layer localization ${\epsilon_1}$,
        the tolerance of accuracy loss ${\epsilon_2}$, and 
        $k$ top items.
	}
	\KwOut{Layer Index, Negative, and Positive Neurons.}

    \tcc{Layer Localization}
    
    $\delta$ $\gets$ $\lambda_l$ $\max \sum_{x \in I} \Delta\big(D_l(z,x),D_l(z',x)\big)$

    $\delta[0]$, $\delta_{max}$ $\gets$ $0.0$, $\lambda_i \max_{j<i} \delta[j]$

    $\rho$ $\gets$ $\lambda_l \frac{\delta[i]-\delta_{max}[i]}{\delta_{max}[i] + \epsilon_1}$

    $l$ $\gets$ $\arg \max_i \rho[i]$

    \tcc{Neuron Localization}
    
    $V_l$ $\gets$ $\lambda_i$ \texttt{stats}$(D_l^i)$ 

    $v_l$ $\gets$ $\lambda_i$ $\lambda_j$ $|\Aa_{\Dd} - \Aa_{\Dd \gets do(l,i,V_l[j])}|\leq \epsilon_2$

    $ACD_l{\gets}{\lambda_j}$ $\eE$ ($k'|$\texttt{do}($l,j,v_l^j{>}0$)) - 
    $\eE$($k'|$\texttt{do}($l,j,v_l^j{=}0$))

    \Return $l$, $top_k(\max ACD_l)$, $top_k(\min ACD_l)$. 

\caption{\textsc{\toolname (Debugging).}}
\label{alg:overall-debugging}
}
\end{algorithm}


%% file: sections/experiments.tex
\section{Experiments}
\label{sec:experiment}

\input{sections/Table_Dataset}
 \noindent \textbf{Datasets and DNN models.} 
We consider $10$ socially critical datasets from the literature of algorithmic fairness.
These datasets and their properties are described in Table~\ref{table:dataset}.
For the DNN model, we used the same architecture as the literature~\cite{zhang2020white,9793943,10.1145/3460319.3464820} and
trained all datasets on a six-layers fully-connected neural network with $\seq{64, 32, 16, 8, 4, 2}$ neurons. We used the same hyperparameters for the all training with num\_epochs, batch\_size, and learning\_rate are set to $1000$, $128$, and $0.01$, respectively.
The accuracy of trained models are reported in Table~\ref{table:bias-mitigation}.

\vspace{0.5em}
\noindent \textbf{Technical Details.} We implemented \toolname with TensorFlow v2.7.0 and scikit-learn v0.22.2.
We run all the experiments on an Ubuntu 20.04.4 LTS OS  sever with AMD Ryzen Threadripper PRO 3955WX 3.9GHz 16-cores X 32 CPU and two NVIDIA GeForce RTX 3090 GPUs. We choose the values $10$, $1000$, $0.025$, $1$, and $1$ for max\_global, max\_local, $\epsilon$, $s\_g$ , and $s\_l$ in Algorithm~\ref{alg:overall-search}, respectively, and take the average of $10$ multiple runs for all experiments. In Algorithm~\ref{alg:overall-debugging}, we used $L_1$-norm, $10^{-7}$, $0.05$, and $3$ for $\Delta$, $\epsilon_1$, $\epsilon_2$, and $k$, respectively. 

\vspace{0.5em}
\noindent \textbf{Research Questions.}
We seek to answer the following three questions using our experimental setup.
\begin{enumerate}[start=1,label={\bfseries RQ\arabic*},leftmargin=3em]

\item  Can \toolname characterize the amounts of information from protected attributes used for the inferences?

\item Is the the proposed search algorithm effective and efficient (vis-a-vis the
state-of-the-art techniques) in generating individual discrimination instances?

\item Can the proposed causal debugging guide us to localize and mitigate the amounts of discrimination?

\begin{tcolorbox}[boxrule=1pt,left=1pt,right=1pt,top=1pt,bottom=1pt]
Our open-source tool \toolname with all experimental subjects are
publicly accessible:
\begin{itemize}
    \item \url{https://github.com/armanunix/Fairness-testing}
\end{itemize}
\end{tcolorbox}

\end{enumerate}
\subsection{Characterizing QID via Search (RQ1)}
\label{sec:rq1}
An important goal is to characterize the amount of information from protected attributes used during the inference of DNN models. Table~\ref{table:bias-magnitude} shows the result of 
experiments to answer this research question. The left side of table shows the initial characteristics such as the number of protected values ($m$), the maximum possible amounts of discrimination ($Q_I$) based on \texttt{min}($\epsilon^{-1},m$), and the initial number of clusters found using samples from the dataset ($K_I$).
The right side of table shows the results after running our search for $1$ hour. The column \#$I$ is the number of QID instances generated, and $K_{F}$ is the maximum number of clusters discovered by \toolname. The column T$_{K_{F}}$ is the time taken to find the maximum number of clusters from an input with initial clusters $K_{I}$ (in seconds). The columns $Q_{\infty}$ and $Q_{1}$
are the quantitative individual discrimination based on min entropy and Shanon entropy, respectively. 
The columns \#$I_{K_{F}^1}$, \#$I_{K_{F}^2}$, and \#$I_{K_{F}^3}$ show the number of test cases with the highest, second-highest, and third-highest QIDs, respectively, that order test cases with their QID severity. 
Overall, the results show that \toolname can find $3.4\times$ more clusters (in average) from the initial characteristics within one minute of search. The DNN for Students dataset showed the largest increase in the number of clusters going from $1.9$ to $10.9$. \toolname found that Adult Income Census dataset has the largest amounts of QID where $4.05$ out of $5.3$ bits ($76.4\%$) from protected variables are used to make decisions. The German Credit dataset with $1.61$ out of $4.0$ bits ($40.0\%$) showed the least amounts of discrimination. For test-case prioritizing, the column \#$I_{K_{F}^1}$ shows our approach to be useful in finding a small percentage of generated test cases with the worst-case discrimination. In $7$ out of $10$ experiments,
\toolname found less than $50$ test cases with severe discrimination out of hundreds of thousands inputs.

\begin{tcolorbox}[boxrule=1pt,left=1pt,right=1pt,top=1pt,bottom=1pt]
\textbf{Answer RQ1:}
The search algorithm is effective in characterizing the
amounts of discrimination via QID. Within $1$ hour, it increased the number of clusters by $3.4\times$ in average, and found instances that used up to
$76\%$ of protected information ($4.05$ out of $5.3$ bits) to infer DNN outcomes.
\toolname is useful to prioritize test cases with their severity where it generates less than $50$ test cases with the maximum QID among hundreds of thousands test cases. 
\end{tcolorbox}


\input{sections/RQ1-Table}

\subsection{Individual Discriminatory Instances (RQ2)}
\label{sec:rq2}
In this section, we compare the efficiency and effectiveness of our search algorithm to the state-of-the-art techniques in searching individual discrimination ($ID$) instances (as defined in Section~\ref{sec:definition}). 
Our baselines are
\textsc{Aequitas}~\cite{udeshi2018automated}, \textsc{ADF}~\cite{zhang2020white}, and \textsc{NeuronFair}~\cite{9793943}.
We obtained the implementations of these tools from their GitHub repositories and configured them according to the prescribed setting to have the best performance. Following these techniques, we report the results for each protected attribute separately.
Table~\ref{table:discriminatory} shows the results of baselines and \toolname in runs of 15 minutes. The results are averaged over $10$ repeated runs. The column \#$ID$ is the total number of generated individual discriminatory instances. The column \texttt{$l\_s$} is the success rate of local stage of searches. We exclude the global success rate since the goal of global phase in our search is to maximize QID whereas the local phase focuses to generate many $ID$ instances.
We calculate success rate as the number of $ID$ found over the total number of generated samples. The columns \texttt{$T.1st$} and  \texttt{$T.1k$} are the amount of time (in seconds) taken to find the first $ID$ instance and to generate $1,000$ individual discriminatory instances (note: $N/A$ in column \texttt{$T.1k$} means that the tool did not generate $1,000$ $IDs$ in the experiment timeout of $900$ seconds in the average of 10 runs).

The result shows that \toolname outperforms the-state-of-the-art in generating many ID instances. In particular, \toolname finds $27.1\times$, $16.0\times$, and $16.0\times$ more $ID$s in the best case
compare to \textsc{Aequitas}, \textsc{ADF}, and \textsc{NeuronFair}, respectively.
\toolname also generates $3.2\times$, $2.3\times$, and $2.6\times$ more $ID$s in the worst case compare to \textsc{Aequitas}, \textsc{ADF}, and \textsc{NeuronFair}, respectively.
The success rate of local search are $20.6\%$, $33.0\%$, $29.6\%$, and $78.2\%$ in average for \textsc{Aequitas}, \textsc{ADF}, \textsc{NeuronFair}, and \toolname, respectively. For the time taken to find the first $ID$, \textsc{Aequitas} achieves the best result with an average of $0.03$ (s) whereas it took \toolname $1.46$ (s) in average to find the first $ID$. In average, \toolname was found to take the lowest time to generate $1000$ $IDs$ with $57.2$ (S), while \textsc{ADF} took $179.1$ (s), \textsc{NeuronFair} took $135.4$ (s), and \textsc{Aequitas} took the longest time at $197.7$ (s).
Overall, our experiments indicate that \toolname is effective in generating $ID$ instances compared to the three state-of-the-art techniques, largely due to the smoothness of the feedback during the local search. 

\begin{tcolorbox}[boxrule=1pt,left=1pt,right=1pt,top=1pt,bottom=1pt]
\textbf{Answer RQ2:}
Our experiments demonstrate that \toolname outperforms the state-of-the-art fairness testing techniques~\cite{udeshi2018automated,zhang2020white,9793943}. In the best case, our approach found $20\times$
more individual discrimination (ID) instances than these techniques with almost $3\times$ more success rates in average.
However, we found that \toolname is slower than those techniques in finding the first ID instance in the order of a few seconds. 
\end{tcolorbox}

\input{sections/RQ2-Table}

\subsection{Causal Debugging of DNNs for Fairness (RQ3)}
\label{sec:rq3}
We perform experiments over the DNN models to study whether
the proposed causal debugging approach is useful in identifying layers and neurons that significantly effect the amounts of discrimination as characterized by $QID$. Table~\ref{table:localization} shows the results of experiments (averaged over $10$ independent runs). The first two columns show the localized layer and its influence (i.e., $l$ and $\rho$ in Algorithm~\ref{alg:overall-debugging}). The next six columns show top $3$ neurons with the positive influence on fairness (i.e., activating those neurons reduce the amounts of discrimination based $Q_{\infty}$). The last six columns show top $3$ neurons with the negative influence on fairness (i.e., activating those neurons increase the amounts of discrimination based $Q_{\infty}$). 
The layer index $2$ is more frequently localized than other layers where the layers $3$, $4$, and $5$ are localized once. 
Overall, the average causal difference (ACD) ranges from $4\%$ to $55\%$ for neurons with positive fairness effects
and from $0.6\%$ to $18.3\%$ for neurons with negative fairness effects. 

Guided by localization, \toolname intervenes to activate neurons with positive fairness influence or de-activate
those with negative influence.
Table~\ref{table:bias-mitigation} shows the results of this mitigation strategy. The columns $A$ and $K$ show the accuracy and the number of clusters (averaged over a set of random test cases) reported by \toolname before mitigation over the DNN model. The columns $A^{=0}$
and $K^{=0}$ are accuracy and the number of clusters reported after mitigating the DNN model by \textit{de-activating} the neuron with the highest negative fairness impacts (as suggested by Neuron$^{-}_{1}$ in Table~\ref{table:localization}). Similarly, the columns $A^{>0}$
and $K^{>0}$ are accuracy and the number of clusters reported after mitigating the DNN model by \textit{activating} the neuron with the highest positive fairness impacts (as suggested by Neuron$^{+}_{1}$ in Table~\ref{table:localization}).
The results indicate that the activation interventions
can reduce QID discrimination by at least $5\%$ with $3\%$ loss of accuracy and up to $64.3\%$ with $2\%$ loss of accuracy.
The de-activation, on the other hand, can improve the fairness by at least $6\%$ with $1\%$ loss of accuracy and up to $27\%$ with $2\%$ loss.

\begin{tcolorbox}[boxrule=1pt,left=1pt,right=1pt,top=1pt,bottom=1pt]
\textbf{Answer RQ3:}
The debugging approach implemented in \toolname identified neurons that 
have at least $5\%$ and up to $55\%$ positive causal effects on the fairness and those which have at least $0.6\%$ and up to $18.3\%$
negative causal effects. A mitigation strategy followed by the localization can reduce the amounts of discrimination by at least $6\%$ and up to $64.3\%$ with less than 5\% loss of accuracy.
\end{tcolorbox}

\input{sections/RQ3-Table-1}
\input{sections/RQ3-Table-2}

%% file: sections/table_dataset.tex
{\footnotesize
\begin{table*}[!t]
\caption{Datasets used in our experiments.}
\centering
\resizebox{0.75\textwidth}{!}{
\begin{tabu}{|l|l|l|ll|l|ll|}
  \hline
  \multirow{2}{*}{\textbf{Dataset}} & \multirow{2}{*}{\textbf{\#Instances}} & \multirow{2}{*}{\textbf{\#Features}} & \multicolumn{2}{c|}{\textbf{Protected Groups}} & \textbf{Num. Protected} & \multicolumn{2}{c|}{\textbf{Outcome Label}} \\
   &  & & \textit{Name} & \textit{Size} & \textbf{Values} (m) & \textit{Label 1} & \textit{Label 0}  \\
  \hline
  Adult & \multirow{3}{*}{$32,561$} & \multirow{3}{*}{$13$} & Sex & 2 & & \multirow{3}{*}{High Income} & \multirow{3}{*}{Low Income} \\ \cline{4-5}
  \textit{Census} &  &    &  Race & 5  & 90 &    & \\ \cline{4-5}
  Income~\cite{Dua:2019} &  &    &  Age & 9  &   &  &  \\
  \hline
    & \multirow{3}{*}{$7,214$} & \multirow{3}{*}{$12$} & Sex & 2 & & \multirow{3}{*}{Did not Reoffend} & \multirow{3}{*}{Reoffend} \\  \cline{4-5}
   \textit{Compas}~\cite{compas-dataset} &  &    &  Race & 2 & 12 &   &    \\ \cline{4-5}
   &  &    &  Age & 3  &  &  &    \\
  \hline
  \textit{German}  & \multirow{2}{*}{$600$}&\multirow{2}{*}{$20$}  &  Sex & 2 & \multirow{2}{*}{16} &\multirow{2}{*}{Good Credit} & \multirow{2}{*}{Bad Credit} \\ \cline{4-5} 
  Credit~\cite{Dua:2019-credit} &  &  & Age & 8 & &  & \\ \cline{4-5} 
  \hline
  \textit{Default} & \multirow{2}{*}{$13,636$}&\multirow{2}{*}{$23$}  &  Sex & 2 & \multirow{2}{*}{12}   &\multirow{2}{*}{Default} & \multirow{2}{*}{Not Default} \\ \cline{4-5} 
  Credit~\cite{Default-Credit}  &  &  & Age & 6 &  &  & \\ \cline{4-5} 
  \hline
  \textit{Heart}  & \multirow{2}{*}{$297$}&\multirow{2}{*}{$13$}  &  Sex & 2 & \multirow{2}{*}{14} &\multirow{2}{*}{Disease} & \multirow{2}{*}{Not Disease} \\ \cline{4-5} 
  Health~\cite{Heart-disease} &  &  & Age & 7 &  & &  \\ \cline{4-5} 
  \hline
  \textit{Bank} Marketing~\cite{Dua:2019-bank}  & $45,211$ & $16$ & Age & 9 & 9 & Subscriber & Non-subscriber \\
  \hline
  \textit{Diabetes}~\cite{Diabetes}  & $768$ & $8$ & Age & 9 & 9 & Positive & Negative \\
  \hline
  \textit{Students}  & \multirow{2}{*}{$1044$}&\multirow{2}{*}{$32$}  &  Sex & 2 & \multirow{2}{*}{16} & \multirow{2}{*}{Pass} & \multirow{2}{*}{Not Pass} \\ \cline{4-5} 
  Performance~\cite{Student-performance} &  &  & Age & 8 &  &  & \\ \cline{4-5} 
  \hline
  \textit{}  & \multirow{3}{*}{$15,830$} & \multirow{3}{*}{$137$} & Age & 9 & & \multirow{3}{*}{Utilized Benefits} & \multirow{3}{*}{Not Utilized Benefits} \\  \cline{4-5}
   MEPS15~\cite{MEP} &  &    &  Race & 2  & 36 &  &    \\ \cline{4-5}
   &  &    &  Sex & 2  &   &  &    \\
  \hline
  \textit{}  & \multirow{3}{*}{$15,675$} & \multirow{3}{*}{$137$} & Age & 9 &  & \multirow{3}{*}{Utilized Benefits} & \multirow{3}{*}{Not Utilized Benefits} \\  \cline{4-5}
   MEPS16~\cite{MEP} &  &    &  Race & 2  & 36  &  &   \\ \cline{4-5}
   &  &    &  Sex & 2  &  &  &    \\
  \hline
\end{tabu}
}
\label{table:dataset}
\end{table*}
}

%% file: sections/RQ1-Table.tex
{\footnotesize
\begin{table*}[ht]
\caption{\toolname characterizes $QID$ for $10$ datasets and DNNs in $1$ hour run (results are the average of $10$ runs).
}
\centering
\resizebox{0.9\textwidth}{!}{%
\begin{tabu}{|l|l|l|l||l|l|l|l|l|l|l|l|}
\hline
\textbf{Dataset} & $m$ & $Q_I$ & $K_{I}$ & \#$I$ & $K_{F}$ & T$_{K_{F}}$ & $Q_{\infty}$ & $Q_{1}$ & \#$I_{K_{F}^1}$ & \#$I_{K_{F}^2}$ & \#$I_{K_{F}^3}$ \\
\hline
Census & $90$ & $5.3$ & $13.54$ & $230,593$ & $35.61$  & $21.04$ & $4.05$ & $2.64$ & $6.0$ & $28.6$ & $111.6$ \\
\hline
Compas & $12$ & $3.6$ & $3.12$ & $157,968$ & $10.24$ & $6.50$ &$1.81$&$1.40$ & $35.2$ & $338.7$ & $1,016.9$\\
\hline
German & $16$ & $4.0$ & $2.34$  & $245,915$ & $9.56$  & $13.14$ & $1.61$ & $1.10$ & $6.6$ & $16.2$ & $54.8$\\
\hline
Default & $12$ & $3.6$ & $5.58$ & $258,105$ & $11.26$  & $10.94$ &$2.10$&$1.78$ & $3,528.8$ & $9,847.2$ & $9,771.0$\\
\hline
Heart & $14$ & $3.8$ & $4.54$ & $270,029$ & $10.01$ & $11.88$ &$2.31$&$1.80$ & $21.7$ & $135.2$ & $579.7$\\
\hline
Bank & $9$ & $3.2$ & $1.45$ & $172,686$  & $8.93$ & $3.68$ &$2.25$&$1.98$ & $5,118.5$ & $13,513.3$ & $20,438$\\
\hline
Diabetes & $10$ & $3.3$ & $2.39$ & $504,414$ & $7.90$ & $0.016$&$1.40$ &$1.11$ & $89.7$ & $609.6$ & $2,310.1$\\
\hline
Students & $16$ & $4$ & $1.90$ & $133,221$ & $10.90$  & $14$&$1.93$ &$1.35$ & $16.0$ & $130.7$ & $128.7$\\
\hline
MEPS15 & $36$ & $5.2$ & $7.03$ & $19,673$ & $18.52$  & $31.62$ &$2.61$&$1.62$ & $2.6$ & $3.5$ & $6.0$\\
\hline
MEPS16 & $36$ &  $5.2$ & $9.06$ & $14,266$ & $19.25$ & $49.16$ &$2.21$&$1.52$ & $2.0$ & $3.5$ & $6.0$ \\
\hline
\end{tabu}
}
\label{table:bias-magnitude}
\end{table*}
}

%% file: sections/RQ2-Table.tex
\begin{table*}[ht]
\caption{Comparison of generating discriminatory instances to the state-of-the-art in $900$ seconds. \textsc{Prot.} is the the protected attributes; \texttt{\#ID} is the number of individual discriminatory instances; $l\_{s}$ is the success rate of the local search; $T.1st$ and $T.1k$ are the time taken to find the first and $1,000^{th}$ discriminatory instance, respectively (in seconds). The results are averaged over $10$ runs of each tool where we report the deviation from the mean in the parenthesis. The best outcomes are highlighted in \textbf{bold} ($k$ is 1,000 and $\epsilon < 0.1$ used for the standard deviation). }
\centering
\resizebox{\textwidth}{!}{%
\begin{tabu}{|l|l|l|l|l|l|l|l|l|l|l|l|l|l|l|l|l|l|}

\hline
\multirow{2}{2.5em}{\textbf{Dataset}} & \multirow{2}{1.5em}{\textbf{Prot.}}  &\multicolumn{4}{| l |}{\textsc{Aequitas}~\cite{udeshi2018automated}} & \multicolumn{4}{| l |}{\textsc{Adf}~\cite{zhang2020white}}  &\multicolumn{4}{| l |} {\textsc{NeuronFair}~\cite{9793943}} & \multicolumn{4}{| l |}{\toolname}\\\cline{3-18}

&&\texttt{$\#ID$} (k)&\texttt{$l\_{s}$} (\%)&\texttt{$T.1$} (s)&\texttt{$T.1k$} (s) &\texttt{$\#ID$} (k)&\texttt{$l\_s$} (\%)&\texttt{$T.1$} (s) &\texttt{$T.1k$} (s) &\texttt{$\#ID$} (k)&\texttt{$l\_s$} (\%)&\texttt{$T.1$} (s) &\texttt{$T.1k$} (s) &\texttt{$\#ID$} (k) &\texttt{$l\_s$} (\%)&\texttt{$T.1$} (s) &\texttt{$T.1k$} (s) \\
\hline
\multirow{3}{2.5em}{Censes} & sex  & 10.4 (1.1) & 10.3 (1.0) & \textbf{0.02 ($\epsilon$)} & 88.5 (13.1) & 18.2 (1.0) & 18.3 (0.1) & 0.5 ($\epsilon$) & 52.1 (7.4) & 21.6 (1.3) & 19.5 (0.8) & 0.05 ($\epsilon$) & 38.8 (6.5)  & \textbf{79.0 (3.2)} & \textbf{74.0 (1.6)} & 0.5 ($\epsilon$) & \textbf{9.8 (0.6)} \\
 & age & 8.8 (7.2) & 29.1 (2.2) & 0.02 ($\epsilon$) & 113.5 (34.8) & 21.6 (1.2) & 55.8 (2.0) & 0.5 ($\epsilon$) & 33.0 (7.3)   & 21.8 (11) & 54.8 (1.8) & \textbf{0.01} ($\epsilon$) &  48.9 (19.7)  & \textbf{112 (2.0)} & \textbf{93.4 (0.6)} & 0.5 ($\epsilon$)  & \textbf{8.0 (1.0)} \\
  & race  & 13.2 (9.7) & 25.2 (1.5) & 0.02 ($\epsilon$) & 89.9 (15.9) & 24.1 (9.0) & 43.3 (1.0) & 0.7 (0.2) & 40.6 (10.0)  & 25.9 (1.1) & 43.5 (2.8) & \textbf{0.01} ($\epsilon$)  &  37.5 (9.9) & \textbf{107 (1.7)} & \textbf{88.1 (0.8)} & 0.5 ($\epsilon$) & \textbf{8.8 (1.3)} \\
\hline
\multirow{3}{2.5em}{Compas} & sex & 12.6 (3.3)  & 13.8 (3.2) & \textbf{0.01} ($\epsilon$) & 118.2 (16.2) & 17.3 (0.7) & 17.4 (0.6) & 0.02 ($\epsilon$) &  49.0 (14.0) & 15.1 (1.1) & 14.3 (0.9)  & 0.02 ($\epsilon$) & 56.3 (9.1)    
 & \textbf{40.2 (1.5)} & \textbf{59.0 (1.4)} & 0.2 ($\epsilon$)  & \textbf{27.6 (5.9)} \\
 & age & 6.8 (0.9) & 10.3 (1.3) & 0.02 ($\epsilon$)  & 150.3 (23.7) & 19.0 (0.9) & 25.6 (1.2) & \textbf{0.01} ($\epsilon$) & 47.6 (5.8) &  14.0 (0.8) & 18.1 (0.9) & 0.01 ($\epsilon$) & 70.1 (15.1) & \textbf{66.2 (3.0)} & \textbf{72.1 (1.6)} & 0.01 ($\epsilon$) & \textbf{19.7 (2.9)} \\
\hline
\multirow{2}{2.5em}{German} & sex & 8.2 (0.8) & 12.7 (1.2) & \textbf{0.01} ($\epsilon$) & 120.7 (20.1) & 14.3 (0.7) & 22.1 (0.8) & 0.02 ($\epsilon$) & 68.7 (12.0) & 12.9 (0.6) & 18.3 (0.8) & 0.09 ($\epsilon$) & 86.4 (14.5) & \textbf{62.6 (1.6)} & \textbf{80.2 (0.6)} & 0.2 (0.2) & \textbf{15.0 (2.1)} \\
 & age & 8.4 (0.6) & 39.8 (2.1) & 0.02 ($\epsilon$) & 118.9 (22.9) & 13.4 (1.1) & 57.4 (3.0) & 0.01 ($\epsilon$) & 60.6 (12.9) & 12.9 (0.8) & 52.5 (2.2) & \textbf{0.01} ($\epsilon$) & 56.5 (12.7) & \textbf{78.7 (0.8)} & \textbf{93.6 (0.2)} & 0.05 ($\epsilon$) & \textbf{11.1 (0.5)} \\
 \hline
\multirow{2}{2.5em}{Default} & sex & 4.6 (0.6) & 5.3 (0.6) & 0.2 (0.2) & 200.0 (45.1) & 12.8 (0.7) & 18.1 (1.1) & \textbf{0.01} ($\epsilon$) & 78.8 (7.3) & 11.4 (0.9) & 14.9 (1.3) & 0.01 ($\epsilon$) & 80.0 (15.4) & \textbf{29.4 (1.5)} & \textbf{38.8 (1.9)} & 2.1 (1.8) & \textbf{22.0 (10.5)} \\
 & age & 7.8 (2.5) &  17.2 (3.2) & 0.08 ($\epsilon$) & 130.4 (73.8) & 13.3 (1.3) & 43.4 (3.1) & \textbf{0.02} ($\epsilon$) & 78.9 (19.8) & 9.0 (1.1) & 29.9 (3.5) & 0.06 ($\epsilon$) & 104.8 (14.7) &  \textbf{50.4 (3.7)} & \textbf{68.4 (2.9)} & 0.6 ($\epsilon$) & \textbf{36.1 (17.1)} \\
\hline
\multirow{2}{2.5em}{Heart } & sex  & 7.9 (3.2) & 9.1 (3.4) & 0.02 ($\epsilon$) & 171.8 (30.5)   & 10.6 (1.0) & 13.6 (1.2) & \textbf{0.01} ($\epsilon$) & 94.9 (16.1) & 10.2 (0.5) & 10.9 (.5) & 0.01 ($\epsilon$) & 84.4 (23.9) & \textbf{84.4 (1.3)} & \textbf{83.9 (1.2)} & 0.04 ($\epsilon$) & \textbf{7.7 (0.4)} \\
 & age  & 16.7 (1.7) & 48.1 (4.0) & 0.02 ($\epsilon$) & 79.0 (13.2) & 25.8 (1.7) & 70.0 (2.5) & \textbf{0.01} ($\epsilon$) & 44.1 (18.6) & 28.7 (1.2) & 68.3 (1.4) & 0.01 ($\epsilon$) & 37.9 (16.5) & \textbf{92.5 (1.6)} & \textbf{95.1 (0.4)} & 0.04 ($\epsilon$) & \textbf{7.5 (0.6)}\\
\hline
Bank  & age   & 11.0 (1.8) & 38.0 (4.1) & 0.04 ($\epsilon$) & 115.0 (42.3) & 7.9 (1.0) & 34.2 (4.2) & \textbf{0.02} ($\epsilon$) & 136 (39.8) & 9.2 (1.1) & 36.6 (4.1) & 0.03 
 ($\epsilon$) & 107.3 (40.2) & \textbf{47.4 (5.6)} & \textbf{87.3 (1.4)} & 0.4 (0.2) & \textbf{25.0 (19.5)} \\
\hline
Diabetes & age  & 15.3 (1.0) & 84.5 (1.7) & \textbf{0.01} ($\epsilon$) & 29.3 (4.7) & 46.8 (0.9) & 76.4 (0.8) & 0.01 ($\epsilon$) & 17.8 (4.2) & 47.9 (1.7) & 76.2 (1.1) & 0.01 ($\epsilon$) & 18.9 (4.7) & \textbf{171 (1.4)} & \textbf{94.2 (0.1)} & 0.04 ($\epsilon$) & \textbf{5.1 (0.2)} \\
\hline
\multirow{2}{2.5em}{Student} & sex  & 2.4 (0.4) & 6.1 (1.0) & 0.03 ($\epsilon$) & 419.5 (58.9)   & 4.7 (0.4) & 12.2 (1.0) & \textbf{0.01} ($\epsilon$) & 200.8 (25.5)  & 4.1 (0.6) & 9.4 (1.3) & 0.02 ($\epsilon$) & 246.5 (43.6) & \textbf{30.7 (1.1)} & \textbf{72.9 (1.3)} & 1.2 (0.7) & \textbf{28.8 (4.0)} \\
 & age  & 1.6 (0.3) & 12.8 (2.3) & 0.04 ($\epsilon$) & 564.8 (0.2k) & 2.7 (0.4) & 24.1 (2.5) & 0.07 ($\epsilon$)  & 330.9 (54.0)    & 2.7 (0.4) & 20.8 (2.8) & \textbf{0.02} ($\epsilon$) & 409.7 (91.5)  & \textbf{43.3 (1.0)} & \textbf{89.3 (0.5)} & 0.5 (0.3) & \textbf{19.6 (1.4)} \\
\hline
\multirow{3}{2.5em}{MEPS15} & sex  & 0.6 (0.1) & 8.0 (1.4) & \textbf{0.01} ($\epsilon$) & N/A & 1.0 (0.2) & 14.7 (2.2) & 0.01 ($\epsilon$) & 337.0 (0.4k)  & 0.9 (0.2) & 11.7 (2.3) & 0.01 ($\epsilon$) & \textbf{68.1 (0.2k)}  & \textbf{6.2 (0.4)} & \textbf{67.9 (2.8)} & 7.2 (6.4) & 174.2 (33.5) \\
 & age & 0.4 (0.1) & 16.1 (1.8) & 0.03 ($\epsilon$) & N/A   & 0.8 (0.2) & 38.3 (6.8) & \textbf{0.02} ($\epsilon$) & \textbf{92.0 (0.3k)} & 0.7 (0.1) & 29.1 (3.5) & 0.05 ($\epsilon$) & N/A & \textbf{6.7 (0.4)} & \textbf{82.1 (2.9)} & 3.2 (3.4) & 136.0 (12.7) \\
  & race  & 0.7 (0.1) & 9.4 (0.8) & \textbf{0.02} ($\epsilon$) & N/A & 1.4 (0.2) & 21.1 (3.4) & 0.02 ($\epsilon$) & 683.3 (0.1k) & 1.0 (0.2) & 14.2 (2.1) & 0.02 ($\epsilon$) & 474.4 (0.4k) & \textbf{6.4 (0.4)} & \textbf{71.0 (2.2)} & 13 (12) & \textbf{146.2 (25.8)} \\
\hline
\multirow{3}{2.5em}{MEPS16} & sex & 0.6 (0.1) & 7.6 (1.2) & 0.02 ($\epsilon$) & N/A  & 1.2 (0.1) & 17.6 (1.9) & \textbf{0.01} ($\epsilon$) & 605.2 (0.3k)  & 1.0 (0.1) & 12.9 (1.8) & 0.01 ($\epsilon$) & 256.0 (0.4k) & \textbf{5.8 (0.4)} & \textbf{74.1 (1.5)} & 0.14 ($\epsilon$) & \textbf{165.3 (19.3)} \\
 & age & 0.3 ($\epsilon$) & 13.9 (2.0) & 0.02 ($\epsilon$) & N/A & 1.0 (0.2) & 42.6 (7.9) & 1.1 (1.1) & 73.8 (0.2k)  & 1.0 (0.2) & 38.1 (5.5) & 0.03 ($\epsilon$) & \textbf{65.5 (0.2k)} & \textbf{6.1 (0.3)} & \textbf{79.3 (1.6)} & 1.4 ($\epsilon$) & 163.7 (12.9) \\
  & race   & 1.1 (0.2) & 14.5 (2.1) & 0.01 ($\epsilon$) & 654.4 (0.3k) & 1.7 (0.2) & 26.3 (3.8) & \textbf{0.01} ($\epsilon$) & 509.3 (0.1k) & 2.1 (0.3) & 27.1 (2.7) & 0.01 ($\epsilon$) & 480.4 (49.1) & \textbf{6.3 (0.4)} & \textbf{76.9 (2.7)} & 0.2 ($\epsilon$) & \textbf{163.6 (20.2)} \\
\hline
\end{tabu}
}
\label{table:discriminatory}
\end{table*}

%% file: sections/RQ3-Table-1.tex
{\footnotesize
\begin{table*}[ht]
\caption{Localization of layers and neurons that causally influence fairness. Neuron$_i^{+}$ shows the index of i-th top neuron that has positive influence on fairness once activated; Neuron$_i^{-}$ shows the index of i-th top neuron that has negative influence on fairness once activated; ACD$_i^{+}$ shows the (normalized) average causal difference of i-th top neuron with positive fairness influence; ACD$_i^{-}$ shows the (normalized) average causal difference of i-th top neuron with negative fairness influence.}
\centering
\resizebox{\textwidth}{!}{%
\begin{tabu}{|l|ll|ll|ll|ll|ll|ll|ll|}
\hline
\textbf{Dataset} & \textbf{Layer Index} & \textbf{Layer Influence} & \textbf{Neuron}$_1^{+}$ & \textbf{ACD}$_1^{+}$ & \textbf{Neuron}$_2^{+}$ & \textbf{ACD}$_2^{+}$ & \textbf{Neuron}$_3^{+}$ & \textbf{ACD}$_3^{+}$ & \textbf{Neuron}$_1^{-}$ & \textbf{ACD}$_1^{-}$ & \textbf{Neuron}$_2^{-}$ & \textbf{ACD}$_2^{-}$ & \textbf{Neuron}$_3^{-}$ & \textbf{ACD}$_3^{-}$  \\
\hline

Census & $2$ & $9.01$ & $N_{19}$ & $0.159$ & $N_{3}$ & $0.150$ & $N_{12}$ & $0.122$ & $N_{15}$ & $0.183$ & $N_{24}$ & $0.163$ & $N_{14}$ & $0.102$ \\

\hline
Compas & $2$ & $20.70$ & $N_{25}$ & $0.516$ & $N_{28}$ & $0.419$ & $N_{7}$ & $0.346$ &$N/A$ & $N/A$& $N/A$ & $N/A$ & $N/A$ & $N/A$ \\

\hline
German & $5$ & $1.79$ & $N_{3}$ & $0.050$ & $N_{1}$ & $0.013$ & $N/A$ & $N/A$ & $N_{0}$ & $0.042$ & $N/A$ & $N/A$ & $N/A$ & $N/A$  \\
\hline
Default & $2$ & $27.58$ & $N_{5}$ & $0.039$ & $N_{27}$ & $0.022$ & $N_{19}$ & $0.014$ & $N_{2}$ & $0.031$ & $N_{13}$ & $0.027$ & $N_{10}$ & $0.019$ \\

\hline
Heart & $4$ & $1.47$ & $N_{5}$ & $0.120$ & $N_{4}$ & $0.048$ & $N_{0}$ & $0.044$ & $N/A$  & $N/A$ &$N/A$  & $N/A$ & $N/A$  & $N/A$  \\

\hline
Bank & $3$ & $6.62$ & $N_{0}$ & $0.495$ & $N_{7}$ & $0.178$ & $N_{1}$ & $0.091$ & $N_{6}$ & $0.057$ & $N_{11}$ & $0.014$ & $N/A$ & $N/A$ \\

\hline
Diabetes & $2$ & $1.67$ & $N_{19}$ & $0.041$ & $N_{6}$ & $0.035$ & $N_{5}$ & $0.031$ & $N_{0}$ & $0.042$ & $N_{29}$ & $0.001$ &  $N/A$  & $N/A$ \\

\hline
Students & $2$ & $4.01$ & $N_{22}$ & $0.550$ & $N_{24}$ & $0.442$ & $N_{8}$ & $0.229$ & $N_{4}$  & $0.084$ & $N_{18}$  & $0.055$ &  $N_{28}$  & $0.026$ \\

\hline
MEPS15 & $2$ & $35.44$ & $N_{24}$ & $0.230$ & $N_{6}$ & $0.167$ & $N_{14}$ & $0.160$ & $N/A$  & $N/A$ &  $N/A$  & $N/A$ &  $N/A$  & $N/A$\\
\hline
MEPS16 & $2$ & $47.46$ & $N_{8}$ & $0.147$ & $N_{11}$ & $0.138$ & $N_{24}$ & $0.114$ & $N_{30}$ & $0.006$ & $N_{30}$  & $0.003$ & $N_{22}$ & $0.001$ \\
\hline
\end{tabu}
}
\label{table:localization}
\end{table*}
}

%% file: sections/RQ3-Table-2.tex
{\footnotesize
\begin{table}[ht]
\caption{$A$ is accuracy, $K$ is the average number of clusters from test cases; $A^{=0}$ is the accuracy after deactivating the neuron with the highest negative
fairness impacts; $K^{=0}$ is the average number of clusters after the deactivation; $A^{>0}$ is the accuracy after activation the neuron with the highest positive fairness impacts; $K^{>0}$ is the average number of clusters after the activation; and $T_I$ is the amount of computation times for localization and mitigation in seconds.}
\centering
\resizebox{0.49\textwidth}{!}{%
\begin{tabu}{|l|l|l|l|l|l|l|l|}
\hline
\textbf{Dataset} & \textbf{A} & \textbf{K} & \textbf{A}$^{=0}$ & \textbf{K}$^{=0}$ & \textbf{A}$^{>0}$ & \textbf{K}$^{>0}$ & \textbf{T}$_I$ \\
\hline
Census & $0.882$ & $22.68$ & $0.85$ & $18.22$ & $0.85$ & $18.68$ & $1,538$\\
\hline
Compas & $0.976$ & $5.19$ & $N/A$ & $N/A$ & $0.971$ & $2.66$ & $1,226$\\
\hline
German & $1.0$ & $5.05$ & $0.992$ & $4.62$ & $0.956$ & $3.53$ & $559$\\
\hline
Default & $0.827$ & $8.08$ & $0.822$ & $6.70$ & $0.826$ & $7.70$ & $1,230$\\
\hline
Heart & $0.96$ & $6.91$ &$N/A$ & $N/A$ & $0.916$ & $6.59$ & $553$\\
\hline
Bank & $0.923$ & $5.95$ & $0.901$ & $4.32$ & $0.893$ & $3.16$ & $858$\\
\hline
Diabetes & $0.993$ & $3.90$ & $0.956$ & $2.91$ & $0.965$ & $3.61$ & $1,238$\\
\hline
Students & $1.0$ & $6.42$ & $0.97$ & $5.79$ & $0.98$ & $2.29$ & $1,206$\\
\hline
MEPS15 & $0.898$ & $9.35$ & $N/A$& $N/A$ & $0.866$ & $6.93$ & $1,348$\\
\hline
MEPS16 & $0.913$ & $7.42$ & $0.906$ & $6.98$ & $0.903$ & $6.27$ & $1,368$\\
\hline
\end{tabu}
}
\label{table:bias-mitigation}
\end{table}
}

%% file: sections/discussion.tex
\section{Discussion}
\label{sec:discuss}
\noindent \textit{Limitation}. In this work, we consider all set of protected values
and perturb them to generate counterfactual. Various perturbations of protected attributes
may yield unrealistic counterfactuals and contribute towards false positives
(an over-approximation of discrimination). This limitation can be mitigated by supplying domain-specific constraints (Age$<$YY$\implies$NOT(married)): we already apply some common-sense constraints
(e.g., to ensure valid range of age). In addition, similar to any dynamic testing methods,
our approach might miss discriminatory inputs and is prone to false negatives. The probability of missing relevant inputs can be contained under a suitable statistical testing (e.g., Bayes factor). In addition,
our debugging approach is similar to pin-pointing suspicious code fragments and is based on causal reasoning of its effect in decision making rather than correlation. But, it is not to furnish explanations or interpretations of black-box DNN functions.

\vspace{0.5em}
\noindent \textit{Threat to Validity}. To address the internal validity and ensure
our finding does not lead to invalid conclusion, we follow established
guideline and take average of repeated experiments. 
To ensure that our results are generalizable and address external validity,
we perform our experiments on $10$ DNN models taken from the literature of fairness testing.
However, it is an open problem whether these datasets and DNN models are sufficiently
representative for fairness testing.

%% file: sections/related-work.tex
\section{Related Work}
\label{sec:related}
\noindent \textbf{Fairness Testing of ML systems.}
\textsc{Themis}~\cite{angell2018themis} presents a causal discrimination notion
where they measure the difference between the fairness metric of two subgroups
by \textit{counterfactual} queries; i.e., they sample individuals with the protected attributes
set to A and compare the outcome to a counterfactual scenario where the protected attributes
are set to B. Symbolic generation (\textsc{SC})~\cite{10.1145/3338906.3338937,agarwal2018automated} 
presents a black-box testing that approximates the ML models with decision trees and
leverage symbolic execution over the tree structure
to find individual discrimination (ID).
\textsc{AEQUITAS}~\cite{udeshi2018automated} uses a two-step approach
that first uniformly at random
samples instances from the input dataset to find a discriminatory instance and then
locally perturb those instances to further generate biased test cases.  
\textsc{ExpGA}~\cite{10.1145/3510003.3510137} proposed a genetic algorithm (GA)
to generate ID instances in natural language processes. The proposed technique
used a prior knowledge graph to guide the perturbation of protected attributes in the NLP tasks.
While these techniques are black-box, they potentially suffer
from the lack of local guidance during the search.
\textsc{ADF}~\cite{zhang2020white} utilized the gradient of the loss function as guidance in generating ID instances.
The global phase explores the input space to find diverse set of individual discrimination
whereas the local phase exploits each instance to generate many individual discriminatory (ID) instances
in their neighborhoods. \textsc{EIDIG}~\cite{10.1145/3460319.3464820} follows similar
ideas to \textsc{ADF}, but uses different computations of gradients.
First, it uses the gradients of output (rather than
loss function) to reduce the computation cost at each iteration. Second, it uses momentum
of gradients in global phase to avoid local optima. 
\textsc{NeuronFair}~\cite{9793943} extends \textsc{ADF} and \textsc{EIDIG} to support
unstructured data (e.g., image, text, speech, etc.) where the protected attributes
might not be well-defined. In addition, \textsc{NeuronFair} is guided by the DNN's
internal neuron states (e.g., the pattern of activation and deactivation) and their
activation difference. Beyond the capability of these techniques, \toolname quantifies the amounts of discrimination,
enables software developers to prioritize test cases, and searches multiple protected attributes at one time.

Beyond the scope of this paper, a body of prior work~\cite{bellamy2019ai,chakraborty2020fairway,DBLP:conf/icse/ZhangH21,10.1145/3468264.3468537,DBLP:conf/icse/Tizpaz-NiariKT022,10.1145/3540250.3549093} considered testing for group fairness. \textsc{Fairway}~\cite{chakraborty2020fairway} mitigates biases
after finding suitable ML algorithm configurations. In doing so, they used 
a multi-objective optimization (\textsc{FLASH})~\cite{8469102}.
\textsc{Parfait-ML}~\cite{DBLP:conf/icse/Tizpaz-NiariKT022} searches the hyperparameter
space of classic ML algorithms via a gray-box evolutionary algorithm to characterize
the magnitude of biases from the hyperparameter configuration. 

\vspace{0.5 em}
\noindent \textbf{Debugging of Deep Neural Network.}
\textsc{Cradle}~\cite{pham2019cradle} traced the execution graph of
a DNN model over two different deep-learning frameworks and used the differences
in the outcomes to localize what backend functions might cause 
a bug. However, since \textsc{Cradle} did not use causal analysis,
it showed a high rate of false positive. \textsc{Audee}~\cite{9286000}
used a similar approach, but it leveraged causal-testing
methods. In particular, it designed strategies to intervene
in the DNN models and tracked how the intervention affected the observed
inconsistencies. We adapted the layer localization of \textsc{Cradle}
and \textsc{Audee}; but our causal localization is developed using
\texttt{do} logic for a meta-property (fairness). \textsc{Audee}
used a simple perturbation of neuron values for 
functional correctness (i.e., any inconsistency shows a bug) without
considering the accuracy or the severity of neuron contributions to a bug.

\vspace{0.5 em}
\noindent \textbf{In-process Mitigation.}
A set of work considers in-process algorithms to mitigate biases in ML
predictions~\cite{zhang2018mitigating,6413831,agarwal2018reductions}. Adversarial debiasing~\cite{zhang2018mitigating} and
Prejudice remover~\cite{6413831} improve fairness by adding constraints to model parameters or the loss function. Exponentiated gradient~\cite{agarwal2018reductions} uses a meta-learning algorithm to infer a family of classifiers that maximizes accuracy and fairness. Different than these approaches, we develop a mitigation approach
that is specialized to handle neural networks for individual fairness. This setting allows us to exploit the layer-based structure of NNs toward causal reasoning and mitigation. We believe that our approach can be extended with in-process mitigation techniques to maximize fairness in the DNN-based decision support systems.

\vspace{0.5 em}
\noindent \textbf{Formal Methods.} We believe that this paper can connect to the rich literature of formal verification and its application. Here, we provide two examples. \textsc{FairSquare}~\cite{10.1145/3133904} certifies
a fair decision-making process in probabilistic programs using a novel verification
technique called the weighted-volume-computation algorithm. \textsc{SFTREE}~\cite{wang2022synthesizing} formulated the problem of inferring fair decision tree as a mixed integer linear programming and apply constraint solvers iteratively to find solutions. 

\vspace{0.5 em}
\noindent \textbf{Fairness in income, wealth, and taxation.} We develop a fairness testing and debugging approach that is uniquely geared toward handling regression problems. Therefore, our approach can be useful to study and address biases in income and wealth distributions~\cite{piketty2003income} among different race and gender. Furthermore, our approach can be useful to study fairness in taxation (e.g., vertical and horizontal equities~\cite{10.1145/3531146.3533204,ICSE-SEIS23}). We left further study in these directions to future work. 

%% file: sections/conclusion.tex
\section{Conclusion}
\label{sec:conclusion}
DNN-based software solutions are increasingly being used in socio-critical applications where a bug in their design may lead to discriminatory behavior.
In this paper, we presented \toolname{}: an information-theoretic model to characterize the amounts of protected information used in DNN-based decision making. 
Our experiments showed that the search and debugging algorithms, based on the quantitative landscape, are effective in discovering and localizing fairness defects. 

%% file: sections/ack.tex
\vspace{1.0 em}
\noindent \textbf{Acknowledgement.}
The authors thank the anonymous ICSE reviewers for their time and
invaluable feedback to improve this paper. 
This research was partially supported by NSF under grant
DGE-2043250 and UTEP College of Engineering under startup package.